\documentclass[
    ,final            
  ]{aipproc}

\layoutstyle{8x11single}

\newcommand{\VmA}{V\!\!\!-\!A}
\begin{document}

\title{Beyond the Standard Model with Precision Nucleon Matrix Elements on the Lattice}

\author{Huey-Wen Lin}{
  address={Department of Physics, University of Washington, Seattle, WA 98195}
}

\begin{abstract}
Precision measurements of nucleons provide constraints on the Standard Model and can discern the signatures predicted for particles beyond the Standard Model (BSM). Knowing the Standard Model inputs to nucleon matrix elements will be necessary to constrain the couplings of dark matter candidates such as the neutralino, to relate the neutron electric dipole moment to the CP-violating theta parameter, or to search for new TeV-scale particles though non-$\VmA$ interactions in neutron beta decay. However, these matrix elements derive from the properties of quantum chromodynamics (QCD) at low energies, where perturbative treatments fail. Using lattice gauge theory, we can nonperturbatively calculate the QCD path integral on a supercomputer.
In this proceeding, I will discuss a few representative areas in which lattice QCD (LQCD) can contribute to the search for BSM physics, emphasizing suppressed operators in neutron decay, and outline prospects for future development.
\end{abstract}
\maketitle


The Standard Model is one of the great successes of modern physics, and all experiments done so far are consistent with the predictions of the Standard Model (SM) within reasonable errors. However, there are certain deficiencies of the Standard Model that lead us to consider physics beyond the Standard Model (BSM). The LHC has been built to discover the mechanism for electroweak symmetry breaking
and to probe certain potential new BSM particles which appear at the TeV scale, such as some candidates for dark matter. Such high-energy colliders in general are well suited to production of high-mass particles directly at their resonances and for probing processes with cross-sections that scale with energy.
However, there are many low-energy precision experiments that also have the ability to probe BSM physics. These generally fall into two classes: experiments that can very precisely measure physics that is precisely predicted by the SM, such as muon $g-2$, the proton weak charge, and flavor-changing processes involving the CKM matrix. Another class of experiments can look for tiny signals in places that the SM says should be either undetectably small or exactly zero, such as the neutron electric dipole moment, or non-$\VmA$ contributions to neutron beta decay.
These low-energy experiments, which usually involves detections through interactions with nuclear matter, will involve the low-energy nonperturbative properties of QCD. Lattice QCD is an ideal theoretical tool to calculate these from the first principles of the Standard Model.
In this proceeding, we give three examples where LQCD can contribute to the search for BSM physics using nucleons.


{\bf The Strange Proton: }
Recent evidence suggests that the composition of dark matter must be cold, with velocities that are non-relativistic, giving it a Compton wavelength above the width of a proton. One promising candidate for dark matter is the supersymmetric particle called the neutralino. If dark matter is neutralinos, they will interact with nuclei at low-energy. A typical value for the cross-section in the Constrained Minimal Supersymmetry Standard Model is about 1 zeptobarn ($10^{-49}\mbox{ m}^2$).
The strange-quark contribution to the proton scalar density and spin are important inputs for spin-independent and -dependent cross-sections. In this proceeding, we will focus on the LQCD calculation of $\sigma_s=m_s\langle N|\bar{s}s| N\rangle$.

\begin{figure}
\includegraphics[height=.18\textheight]{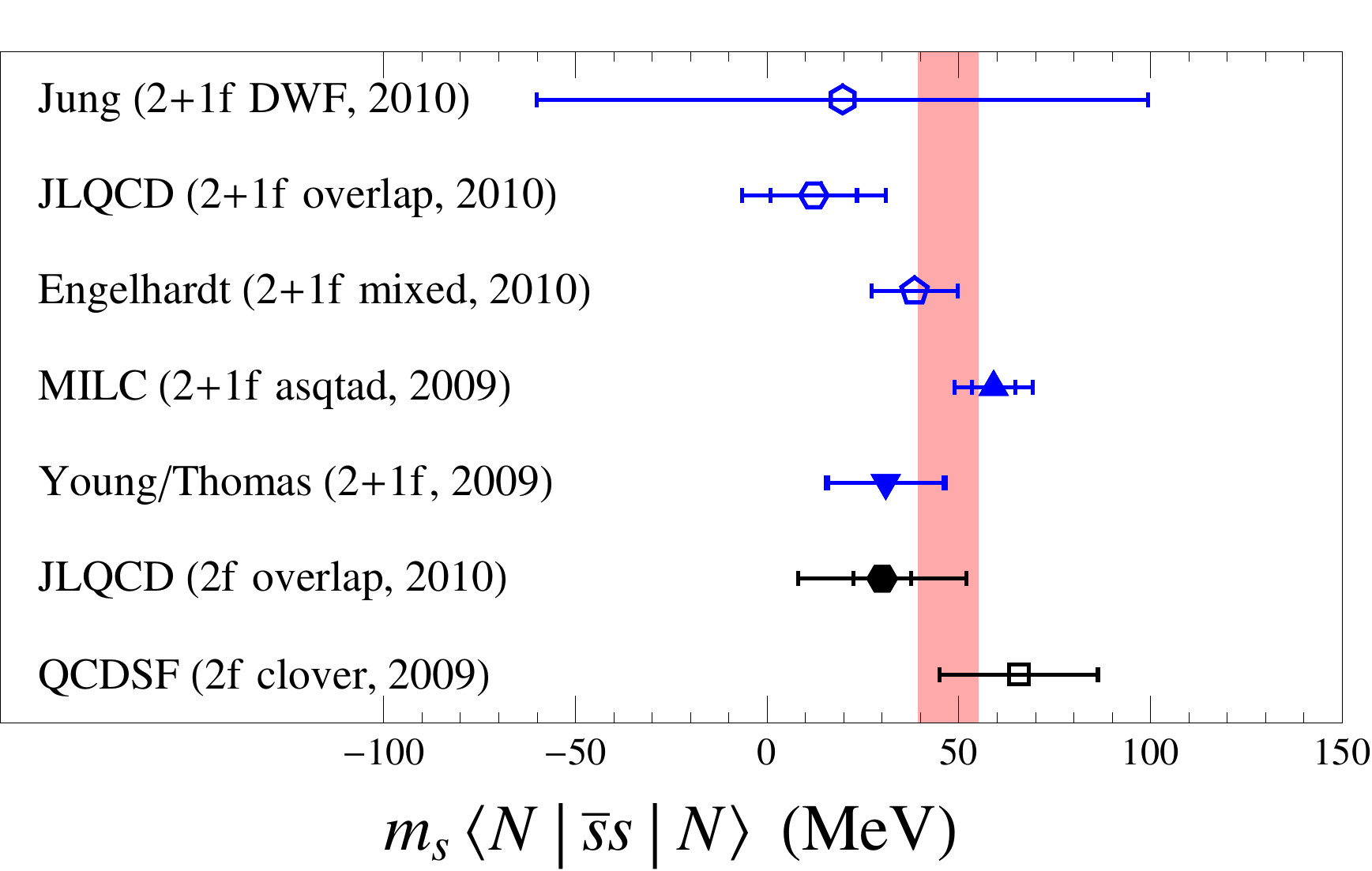}
\includegraphics[height=.18\textheight]{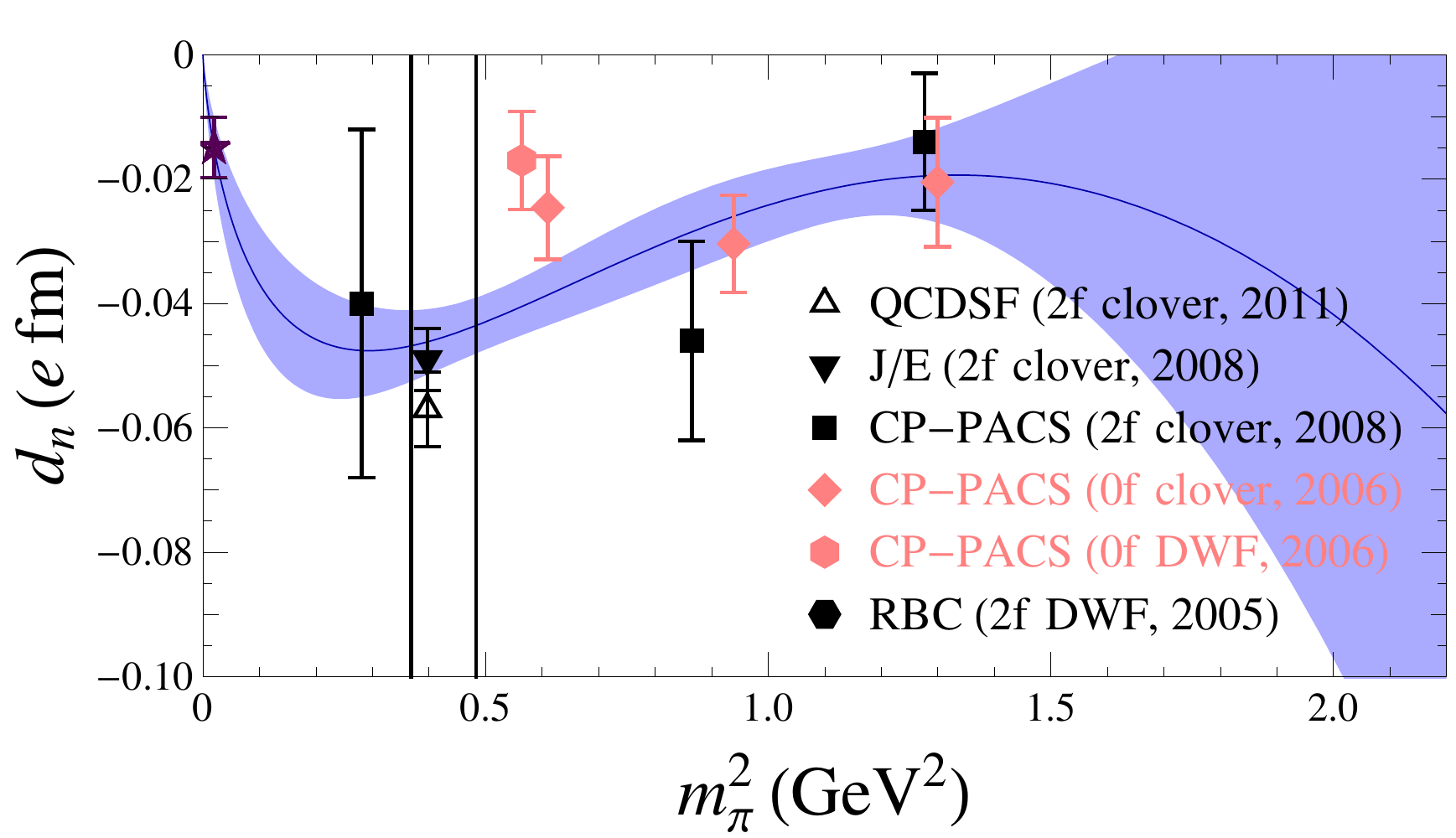}
\caption{\label{fig:fig1}
(left) Summary of dynamical LQCD calculations of the nucleon sigma term $m_s\langle \overline{s}s\rangle$ in MeV. The (pink) band indicates the global average after weighting individual points according to their systematics.
(right) Summary of the latest dynamical $d_n$ calculations as a function of $m_\pi^2$.
The band takes all the lattice points with weighted errobar and globally extrapolates the various calculations to the physical pion mass point (the leftmost star point).
}
\end{figure}

One can directly calculate the matrix element which involves strange-quark loops and proton correlators. Unfortunately, such a direct calculation is noisy, so various techniques have been developed to improve the signal (e.g. JLQCD, QCDSF, BU, Engelhardt)\cite{strange:2011a}.
Alternatively, one could use the Feynman-Hellman Theorem by studying how the nucleon mass varies with the strange-quark mass, taking the numerical derivative $dM_N/dm_s$ either by direct SU(3) fitting to the baryon masses (e.g. Young/Thomas) or by reweighting strange part of action (e.g. Jung, MILC)\cite{strange:2011b}.
The values of $\sigma_s$ from various calculations are shown on the left-hand side of Fig.~\ref{fig:fig1}.
We perform a global fit to current dynamical lattice data (weighted by lattice spacing, lightest $m_\pi$, dynamical strange and other quality factors) and obtain $\sigma_s^{\rm LQCD}\!\!\!\!=47(8)$~MeV, which is consistent with the value obtained using only two lattice inputs in Ref.~\cite{Giedt:2009mr}. The cross-section constraints from various models in the CMSSM can be found in Ref.~\cite{Giedt:2009mr}.

{\bf Neutron Electric Dipole Moment: }
The neutron electric dipole moment is a measure of the distribution of positive and negative charge inside the neutron. To generate a finite nEDM, one needs processes that violate CP-symmetry, for example by adding a CP-odd $\theta$-term to the lagrangian. The SM value of this quantity is very small, $10^{-30}$~$e\cdot\mbox{cm}$. Although experiments do not have the necessary precision to measure the SM value; many BSM models predict values that are higher than the experimental upper bounds, 
including some parts of the parameter space for certain SUSY models.

There are three main approaches to extracting the nEDM using LQCD. One is a direct measurement, by studying the electromagnetic form factor $F_3$ under the new QCD lagrangian including the CP-odd term (RBC, J/E, CP-PACS 2005). This requires us to extrapolate the form factors to $q^2=0$, which can introduce systematic error and exacerbate the statistical error.
Another method is through indirect measurement by
looking at the energy difference for two different spin states of the nucleon at zero momentum under an external static and uniform electric field (CP-PACS 2008).
Or finally, one can measure the product of the anomalous magnetic moment of neutron $F_2(q^2=0)$ and $\tan(2\alpha)$ (QCDSF)
in a CP-violating system.
The collected dynamical LQCD results\cite{nedm:2011} 
on $d_n$ are shown on the right-hand side of Fig.~\ref{fig:fig1};
when combining all data and extrapolating to the physical pion mass, we obtain
$d_n^{\rm LQCD}\!\!\!\!=-0.015(5)$~$\theta e\cdot\mbox{fm}$. 
More updated and precise calculations from various groups are currently in progress.

{\bf Non-$\VmA$ Interactions in Neutron Beta Decay: }
A third example of probing BSM with nucleons is to look for signatures of contributions due to scalar or tensor interactions in neutron beta decay\cite{pndme:2011}. 
The notion is quite similar to how Fermi theory led to the discovery of the electroweak interaction and its bosons.
Beta decay was originally explained by Fermi by adding a new term to the fundamental lagrangian describing 4-fermion interactions. Such a theory introduces a coupling, the Fermi constant, that has units of inverse-energy squared, with an energy scale around 100~GeV.
As it turns out, the Fermi theory is a low-energy effective theory approximating the electroweak theory, which has 3 vector bosons, the W's and Z. The theory was later probed in high-energy proton-antiproton experiments at CERN, and the new particles were found with resonances around the scale predicted by this interpretation of the Fermi theory.
We can imagine that new particles beyond the Standard Model can be predicted in just such a way: by first detecting the low-energy effective operators and later directly probing them in a high-energy experiment.

The new BSM physics will enter the neutron beta-decay hamiltonian,
$\varepsilon_i^{\rm BSM} \hat{O}_i^{\rm lept} \times \hat{O}_i^{\rm quark}$,
with the coefficients $\varepsilon={v}^2/(\Lambda_i^{\rm BSM})^2$, where $\Lambda_i^{\rm BSM}$ are related to the TeV scale of the particles. The leptonic part is understandable using analytic techniques, but the quark operator in the context of the nucleon will introduce some unknown coupling constants,
$g_T = \langle p | \overline{u}\sigma_{\mu\nu} d | n \rangle$ and $g_S = \langle p | \overline{u} d | n \rangle$, which are nonperturbative functions of the nucleon structure, described in the SM by QCD.
Any deviation from the SM $\VmA$ current coming from the new scalar and tensor interactions in the effective theory will require knowledge of the couplings $g_S$ and $g_T$ to understand.

\begin{figure}
\includegraphics[width=.42\textwidth]{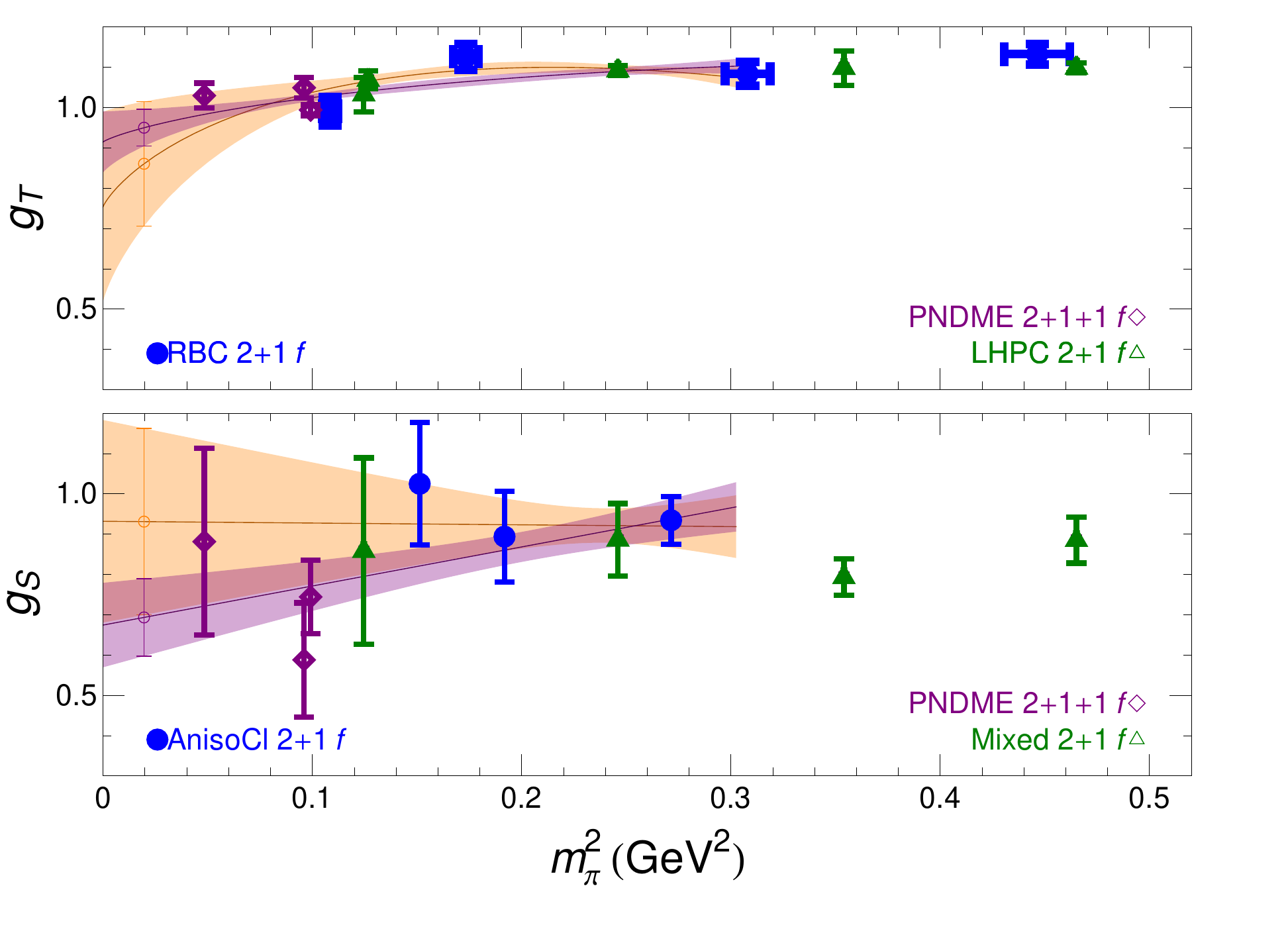}
\includegraphics[width=.32\textwidth]{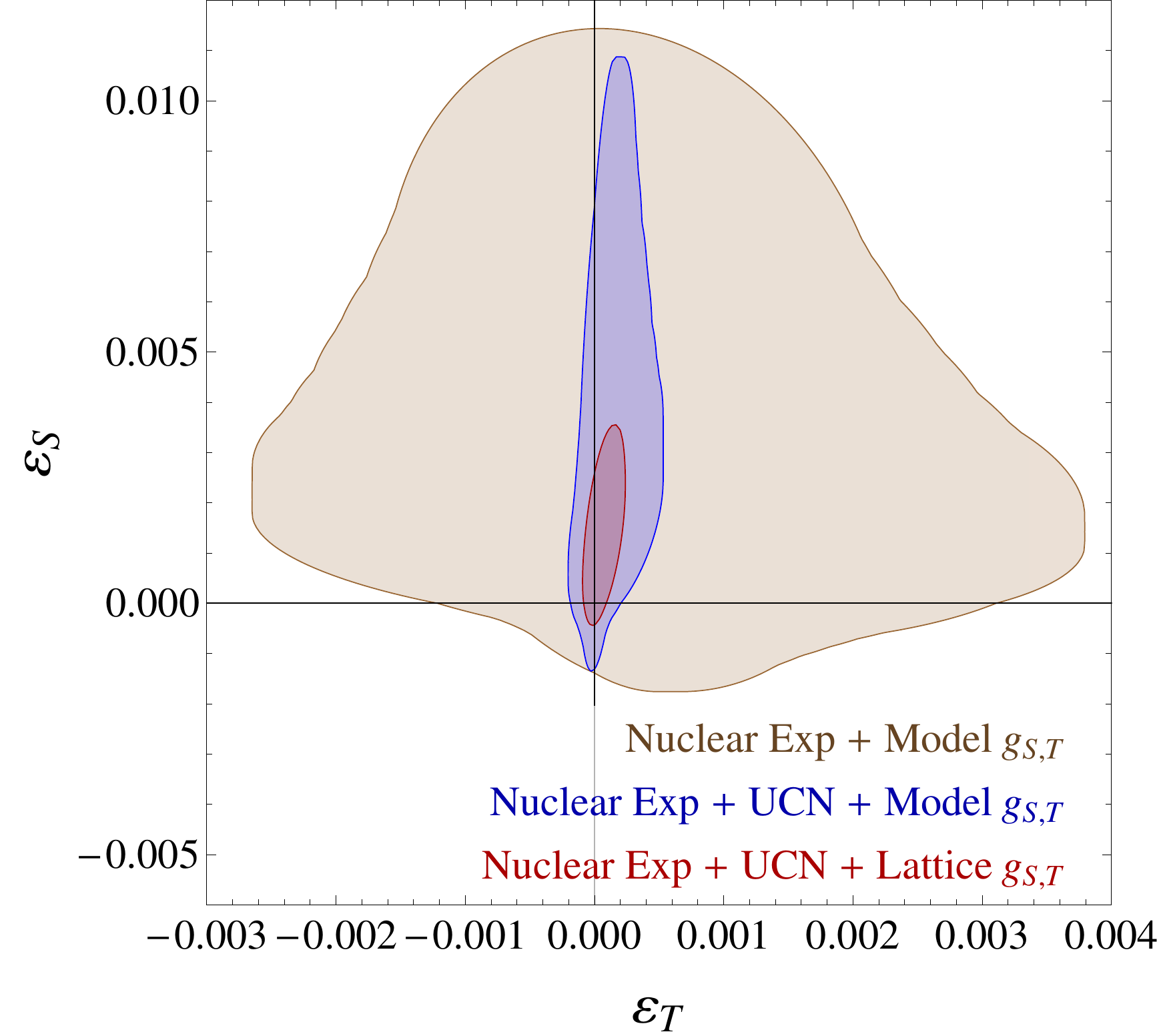}
\caption{\label{fig:fig2}
(left) Global analysis of all $N_f=2+1$ lattice calculations of $g_T$ and $g_S$. The left-most points are the extrapolated values at the physical pion mass. The two bands show different extrapolations in the pion mass: with and without the lighter PNDME collaboration points; the final uncertainty is significantly improved due to the chiral extrapolation.
(right) $\epsilon_{S}$-$\epsilon_{T}$ allowed parameter region using different experimental and theoretical inputs.
}
\end{figure}

The tensor charge ($g_T$) is also known as the zeroth moment of transversity;
experimentally, one can extract the contribution from individual quarks as a function of the quark momentum fraction $x$, at certain $Q^2$ through processes such as semi-inclusive deep inelastic scattering.
To obtain $g_T$, one need to integrate over all possible $x$, that is, from 0 to 1, through an interpolation of the few available $x$ at each $Q^2$.
After combining a few experiments, the best experimental estimate (using model inputs) for this quantity is obtained at $Q^2=0.8\mbox{ GeV}^2$ (instead of $Q^2=0$ as in the correct definition of $g_T$): $0.77^{+0.18}_{-0.24}$ with combined uncertainty around 25\%.
There are also theoretical models that attempt to make estimates of the tensor charge, such as Nambu--Jona-Lasinio model and the chiral quark soliton model; unfortunately, they are not consistent with each other. And using QCD sum rules, still yields a large uncertainty.
The other important ingredient for the new interactions is the scalar charge $g_S$.
There are no experimental measurements of this quantity, and the theoretical estimations (from different model approximations) give rather loose bounds to the quantity: $0.25 \leq g_S \leq 1$.

Lattice calculation of $g_T$ and $g_S$ is rather straightforward.
The tensor charge has been studied a few times in the past using $N_f=2+1$ dynamical ensembles in LQCD,
while $g_S$  has not been studied until now; a summarized LQCD results are shown in the left-hand side of Fig.~\ref{fig:fig2}.
All of these results, by different collaborations, using different fermions actions, are generally in good agreement, although the error bars shown here contain only statistical error.
We further globally analyze all the available lattice data, including the lightest available from PNDME, around 220~MeV, and obtain $g_T^{\rm LQCD}= 0.95(5)$ and $g_S^{\rm LQCD}= 0.69(9)$ through chiral extrapolation to the physical pion mass.

Finally, we can combine the low-energy tensor and scalar charges with experimental data to determine the allowed region of the epsilon parameter space.
We can combine the current knowledge of $g_{S,T}$ and the existing experimental data for the $0^+ \rightarrow 0^+$ transition and nuclear beta decay
(such as $\beta$ symmetry in Gamow-Teller $^{60}\mbox{Co}$,
longitudinal polarization ratio between Fermi and Gamow-Teller transitions in $^{114}\mbox{In}$,
positron polarization in polarized $^{107}\mbox{In}$
and beta-neutrino ccorrelation parameters in nuclear transitions).
Using the $g_{S,T}$ from the model estimations and combining with the existing nuclear experimental data, we get the constraints shown the outermost band on the lower-right of Fig.~\ref{fig:fig2}. There is an on-going ultra-cold neutron experiment studying neutron beta decay at LANL to look for deviations from the SM in
the Fierz term and the neutrino asymmetry parameter %
to the level of $10^{-3}$ by 2013. Combining those expected data and existing measurements, and again, using the model inputs of $g_{S,T}$, we see the uncertainties in $\epsilon_{S,T}$ are significantly improve.
This shows the importance of the precision experimental inputs.
Finally, using our present LQCD values of the scalar and tensor charges, combined with the expected 2013 precision of experimental bounds
we found the constraints on $\epsilon_{S,T}$ are further greatly improved, shown as the innermost region.
This corresponds to lower bounds for the scale $\Lambda_{S,T}$ at 2.9 and 11.2~TeV, respectively, for new physics in these channels.




\begin{thebibliography}{17}
\expandafter\ifx\csname natexlab\endcsname\relax\def\natexlab#1{#1}\fi
\providecommand{\enquote}[1]{``#1''}
\expandafter\ifx\csname url\endcsname\relax
  \def\url#1{\texttt{#1}}\fi
\expandafter\ifx\csname urlprefix\endcsname\relax\def\urlprefix{URL }\fi
\providecommand{\eprint}[2][]{\url{#2}}

\vspace{-0.3cm}
\bibitem[strange (2011)]{strange:2011a}
H.~Ohki, et~al., \emph{Phys. Rev.} \textbf{D78}, 054502 (2008);
G.~Bali, et~al., \emph{PoS} \textbf{LAT2009}, 149 (2009);
R.~Babich, et~al.  (2010), \eprint{1012.0562};
K.~Takeda, et~al., \emph{Phys. Rev.} \textbf{D83}, 114506 (2011);
S.~Collins, et~al., \emph{PoS} \textbf{LATTICE2010}, 134 (2010);
K.~Takeda, et~al., \emph{PoS} \textbf{LATTICE2010}, 160 (2010);
M.~Engelhardt, \emph{PoS} \textbf{LATTICE2010}, 137 (2010).

\bibitem[strange (2011)]{strange:2011b}
D.~Toussaint, and W.~Freeman, \emph{Phys. Rev. Lett.} \textbf{103}, 122002
  (2009);
R.~D. Young, and A.~W. Thomas, \emph{Phys. Rev.} \textbf{D81}, 014503 (2010);
C.~Jung, \emph{LAT2010}  (2010).

\bibitem[Giedt et~al.(2009)]{Giedt:2009mr}
J.~Giedt, A.~W. Thomas, and R.~D. Young, \emph{Phys. Rev. Lett.} \textbf{103},
  201802 (2009), \eprint{0907.4177}.

\bibitem[nedm (2011)]{nedm:2011}
F.~Berruto, et~al., \emph{Phys. Rev.} \textbf{D73},
  054509 (2006);
E.~Shintani, et~al., \emph{Phys. Rev.} \textbf{D72}, 014504 (2005);
E.~Shintani, et~al., \emph{Phys. Rev.} \textbf{D75}, 034507 (2007);
S.~Aoki, et~al.  (2008), \eprint{0808.1428};
E.~Shintani, S.~Aoki, and Y.~Kuramashi, \emph{Phys. Rev.} \textbf{D78}, 014503
  (2008);
G.~Schierholz, \emph{LAT2011}  (2011).

\bibitem[pndme (2011)]{pndme:2011}
T.~Bhattacharya, et~al., \emph{in preparation}, (2011).

\end{thebibliography}
\end{document}